\documentclass[letter]{aa} 
\usepackage{graphicx}
\usepackage{txfonts}
\usepackage[]{natbib}
\usepackage{multirow}
\usepackage[colorlinks=true,linkcolor=blue,citecolor=blue]{hyperref}
\def\deg{$^{\circ}$~}
\def\degb{^{\circ}}

\begin{document}

 \title{Independent confirmation of $\beta$ Pictoris b imaging with NICI\thanks{Based on data retrieved from the Gemini archive}}

 \author{
 	A. Boccaletti\inst{1},  
	A.-M. Lagrange\inst{2}, 
	M. Bonnefoy\inst{3}, 
	R. Galicher\inst{1},
	G. Chauvin\inst{2}
 }
 
 \offprints{A. Boccaletti, \email{anthony.boccaletti@obspm.fr} }

 \institute{LESIA, Observatoire de Paris, CNRS, University Pierre et Marie Curie Paris 6 and University Denis Diderot Paris 7, 5 place Jules Janssen, 92195 Meudon, France
             \and
             Institut de Plan\'etologie et d'Astrophysique de Grenoble, Universit\'e Joseph Fourier, CNRS, BP 53, 38041 Grenoble, France
             \and
	   Max Planck Institute for Astronomy, K\"{o}nigstuhl 17, D-69117 Heidelberg, Germany
                   }
   \date{Received 7 February 2013; Accepted 21 February 2013}


\authorrunning{A. Boccaletti et al.}
\titlerunning{betaPic NICI}

 \abstract
{$\beta$ Pictoris b is one of the most studied objects nowadays since it was identified with VLT/NaCo as a bona-fide exoplanet with a mass of about 9 times that of Jupiter at an orbital separation of 8-9\,AU. The link between the planet and the dusty disk is unambiguously attested and this system provides an opportunity to study the disk/planet interactions and to constrain formation and evolutionary models of gas giant planets. 
Still, $\beta$ Pictoris b had never been confirmed with other telescopes so far.
}
{
We aimed at an independent confirmation using a different instrument.
}
{We retrieved archive images from Gemini South obtained with the instrument NICI, which is designed for high contrast imaging. The observations combine 
coronagraphy and angular differential imaging and were obtained at three epochs in Nov. 2008, Dec. 2009 and Dec. 2010.}
{We report the detection with NICI of the planet $\beta$ Pictoris b in Dec. 2010 images at a separation of $404\pm10$\,mas and $PA=212.1\pm0.7\degb$. It is the first time this planet is observed with a telescope different than the VLT. }
{}

 \maketitle

%

\section{Introduction}

The A-type star $\beta$ Pictoris is the subject of many attention since the discovery of a planet orbiting around \citep{Lagrange2009a}. 
The first observation from the Very Large Telescope in 2003 reported a candidate companion aligned with the debris disk at a separation of $\sim$0.41" to the North-East (PA = 34.4$\degb$).
The candidate planet, undetected in early 2009 \citep{Lagrange2009b}, was finally recovered at $\sim$0.3" to the South-West at the end of the same year \citep{Lagrange2010} definitely confirming its companionship with the star and semi-major axis smaller than 15\,AU. 
Given the assumption on the age of this system ($\sim$12\,Myr) and its distance (19.44\,pc), the candidate planet was attributed a mass of about 9\,M$_{Jup}$ based on evolutionary models \citep{Baraffe2003} and later confirmed with more detailed photometric works \citep{Quanz2010, Bonnefoy2011, Bonnefoy2013} and upper limits from radial velocity \citep{Lagrange2012a}. {  Owing} to systematic follow-up observations the orbit was characterized by \citet{Chauvin2012} who determined a semi-major axis of 8-9\,AU with low eccentricity ($<$0.17). 
\citet{Lagrange2012b} demonstrated that the planet is inclined with respect to the main disk 
and therefore likely responsible for the observed warp of planetesimals \citep{Mouillet1997, Heap2000}. 
Recently, a global analysis of the planet near-IR photometry and careful comparison to models were presented in \citet{Bonnefoy2013}.
Moreover, $\beta$ Pictoris b,  the closest planet to its parent star ever imaged so far, may well have been formed via core accretion. 

So far, any observations of $\beta$ Pictoris b reported in the literature were obtained with NaCo (Nasmyth Adaptive Optics System and Near-Infrared Imager and Spectrograph), the near-IR AO-assisted camera at the Very Large Telescope \citep{Rousset2003, Lenzen2003}. A single tentative observation at Keck was presented in \citet{Fitzgerald2009} but was performed at a time where the planet was angularly too close to the star for being detectable and at relatively high airmass. 
In this paper, we present the first analysis of data collected at Gemini South and retrieved from the {  Gemini} archive.

\section{Observations}

\begin{table*}[ht]
\caption{  Log of observations which indicates for each data set the date of observation (col. 2), the mask radius (col. 3), the filters (col. 4), the UT at start and at end of the sequence (col. 5), the exposure time of individual frame (col. 6), the amplitude of the parallactic angle variation (col. 7) and the total exposure time integrated (col. 8). }
\begin{center}
\begin{tabular}{ccccccccccc} \hline\hline
Set 	&	Date			&  	Mask 	& Filters 			&	UT start/end		&	Exp. time  & 	Parallactic angle  & Total exp. time	\\	
	&				&	[arcsec]		&				&					&	[s]		& amplitude [deg]	& [s]	\\	
\hline \\
1	&	2008.11.22	&	0.22		& CH4L / CH4S		&	07:23:40 / 08:45:15	&	3.04		&	68.6 	&  5411	\\
\hline \\
2	&	2009.12.02	&	0.32		& CH4L / CH4S		&	05:21:10 / 06:28:11	& 	4.18 / 2.66&	39.7	&  3224	\\
3	&	2009.12.03	&	0.32		& CH4L / CH4S		&	05:58:58 / 08:51:50	& 	3.04		&	67.5	&  8086 	\\
4	&	2009.12.04	&	0.32		& - / H				&	05:46:38 /  08:48:19 & 	9.88		&	72.1	&  7884	\\
\hline \\
5	&	2010.12.25	&	0.22		& Ks / H			&	04:24:03 / 05:05:37	&	1.14		&	28.7	& 684	\\
6	&	2010.12.25	&	0.22		& Ks / H			&	05:15:01 / 07:20:34	&	0.76		&	43.1	& 3268	\\ 
7	&	2010.12.26	&	0.22		& CH4L / J		&	06:18:42 / 06:51:54	&	0.76		& 	9.7	& 1064 	\\
8	&	2010.12.26	&	0.22		& CH4L / J		&	06:59:19 / 07:40:54	&	1.14 		&	8.5	& 1425 	\\

\end{tabular}
\end{center}
\label{tab:log}
\end{table*}

The Near Infrared Coronagraphic Imager \citep[][NICI]{Toomey2003}, installed at  the 8-m telescope, Gemini South, is designed to take advantage of the Spectral Differential Imaging (SDI) for exoplanets detection \citep{Racine1999}. The instrument is based on a near-IR (1-5$\muup$m) dual-band imager in which two images are formed simultaneously (owing to a beam splitter) on two separate detectors. 
The two spectral channels are composed with the same filters so that it is possible to combine either two spectrally adjacent filters (CH4L with $\lambda=1.578\muup$m and CH4S with $\lambda=1.652\muup$m) for methanated-object detection or any near IR broad-band filters. The pixel scale is 17.932 and 17.970\,mas per pixel (as measured for each detector in Nov. 2010 at an accuracy of 0.1\%) hence providing a field of view of about 18.4". NICI also allows for Angular Differential Imaging \citep[][ADI]{Marois2006} in order to improve the final contrast. 

As for the coronagraphic part, several semi-transparent Lyot masks (radii ranging from 0.22" to 0.90") are used in combination with several pupil stops undersized with respect to the telescope pupil (outer edge stopped from 80 to 95\% that of the pupil). In the following we will refer to two coronagraphic configurations that associate the 0.22" and 0.32" masks to the 95\% pupil stop. The focal masks are transmitting {  a fraction  of the light ($\sim$1:300 for the 0.32" mask)} at the center and have a flat-topped Gaussian profile. The transmission is therefore not uniform in the field near the mask and has to be taken into account when it comes to the photometry of faint companions. Each mask is deposited on a substrate which carries its own imperfections. 

While SDI and ADI were already implemented on large telescopes, NICI is certainly the first instrument fully designed for this specific purpose \citep{Liu2010} and has an important role to play in preparing the observations with the near to come planet finders namely SPHERE \citep{Beuzit2008} and GPI \citep{Macintosh2008}. 

We retrieved three epochs of observations for $\beta$ Pictoris from the NICI archive released on Nov. 24th, 2012, corresponding to programs GS-2008B-SV-1, GS-2009B-Q-500, GS-2010B-Q-501. The list of available data is given in Tab. \ref{tab:log}.

Data reduction was performed in a similar way as for NaCo data using our own pipeline \citep{Bonnefoy2013, Boccaletti2012} and we followed the guidelines presented in other papers reporting NICI observations \citep{Biller2010, Wahhaj2011, Nielsen2012}. Here, the two channels are considered separately. The images are dark-subtracted and flat fielded. We did not correct for distortion since we are interested in a very small field around the star (the planet is expected at a separation closer than 0.4" at these epochs) and we did not attempt to perform SDI processing when the CH4L/CH4S filters are in use. Flat fields are usually not available for the dates of observation so we used calibrations taken several months apart. 
This prevents to account for inhomogeneities in the substrate of the focal masks as long as we noticed a drift {  of the mask position by} several pixels between the flat fields and the science data. In addition, background was not subtracted out since we are lacking of sky observations. 

Conversely to NaCo observations of $\beta$ Pictoris the data are not saved in data cubes but summed, hence totaling long exposures of the order of 40-60s. While this is not critical with respect to the field rotation at the separation of the planet it can be an issue for selecting the best frames. 
Since the observations were obtained with a semi-transparent focal Lyot mask, the selection is based on the frame with the minimal amount of intensity (integrated over all the pixels) indicative of a better centering or a good AO correction. For a given sequence of frames, we measured the total flux and retained the frames for which the flux is lower than three times the standard deviation of the total flux above the minimal flux. For the data presented hereafter the selection retained more than 80\% of the frames. 
 
As for the registration of the coronagraphic images we proceeded in two steps. First, we applied a rough estimation of the star position with a gaussian fit of the images thresholded at typically a few percent of the maximum flux, as we did in \citet{Boccaletti2012}. 
Then, we performed a more precise determination of the star location with Moffat fitting of the central attenuated spot behind the semi-transparent Lyot mask which is known to move linearly with the actual star position \citep{Lloyd2005}. The coronagraph centering is relatively good in 2008 and 2009 data but features a noticeable drift  in 2010 data which amounts to several pixels (the 0.22" mask is about 12 pixels in radius {  for comparison}).
The parallactic angle is calculated from the azimuth and declination of the telescope pointing contained in the FITS headers. 

Out of mask unsaturated images of the star used to determine the PSF shape and intensity are not available in the archive. To mimic a PSF we used the central spot behind the mask {  as proposed in the NICI documentation\footnote{http://www.gemini.edu/sciops/instruments/nici/imaging}}. We measured a FWHM of 3.6 pixels equivalent to 64 mas in the Ks band. However, this spot is not rigorously an attenuated copy of the star since the energy  in the center results from the addition of the coronagraphic diffraction and the semi-transparent mask transmission. Hence, the photometric information provided by this signal is difficult to calibrate as it depends on the mask centering, the AO correction, the mask size and the wavelength.

Then, we processed the data with a set of ADI algorithms: cADI, rADI \citep{Marois2006}, LOCI \citep{Lafreniere2007} and also KLIP which makes use of a Principal Component Analysis \citep{Soummer2012}.  For a brief description of these algorithms and relevant control parameters see \citet{Lagrange2012b, Boccaletti2012}. Several parameters for these algorithms were tested to achieve the best detection performance (separation criteria are indicated in Tab. \ref{tab:photom}).

\begin{figure}[t]
\centerline{
\includegraphics[width=4.5cm]{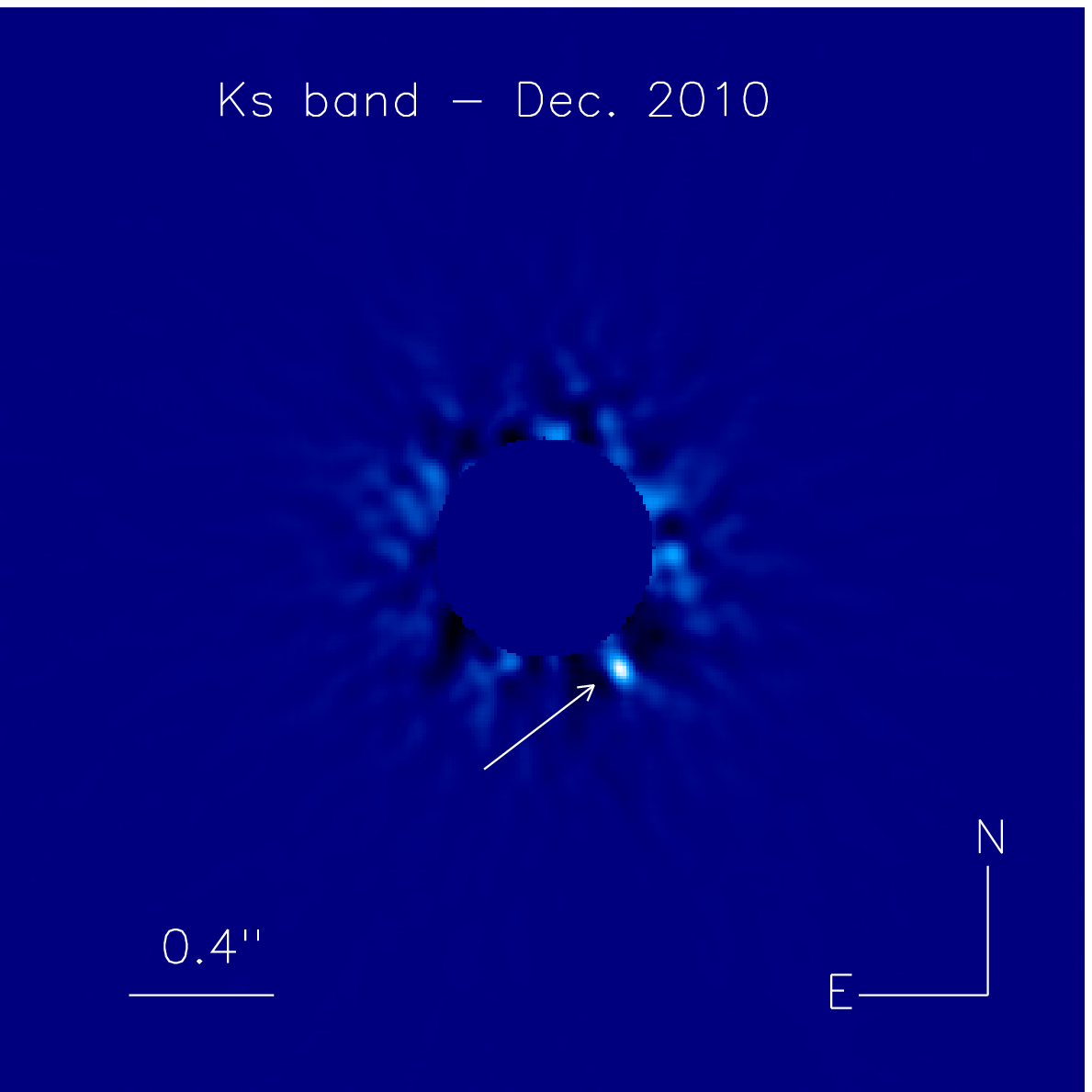}
\includegraphics[width=4.5cm]{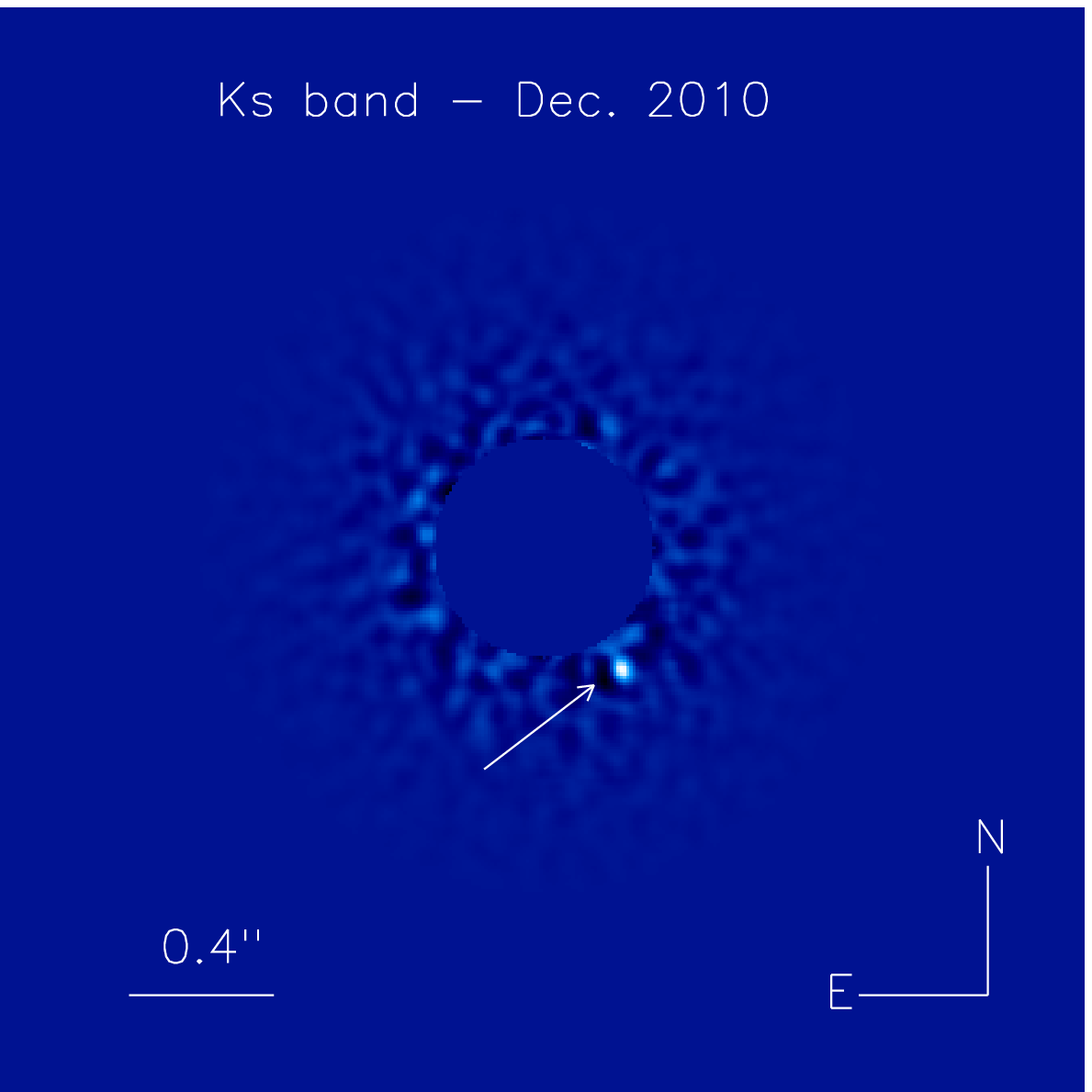}}

\centerline{
\includegraphics[width=4.5cm]{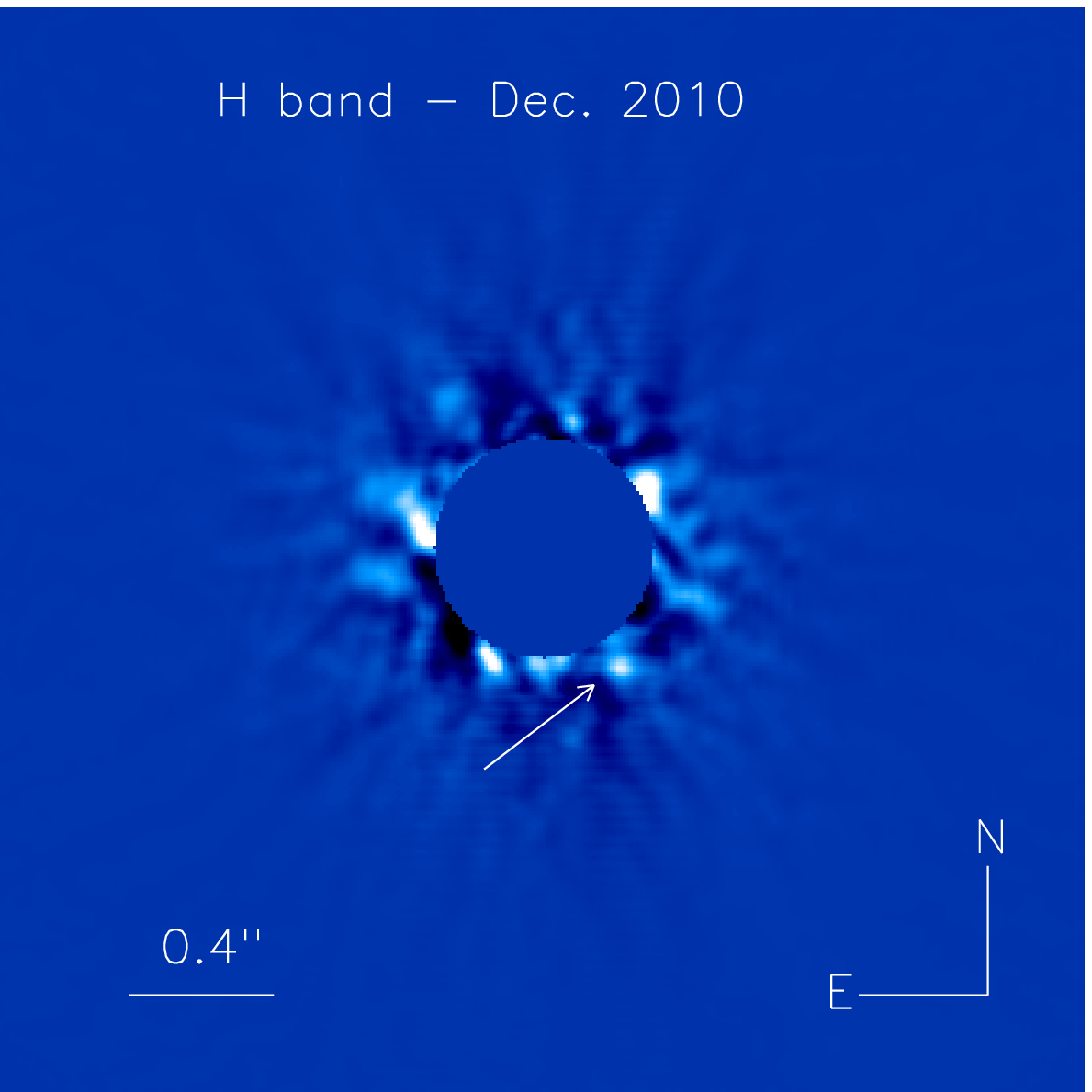}
\includegraphics[width=4.5cm]{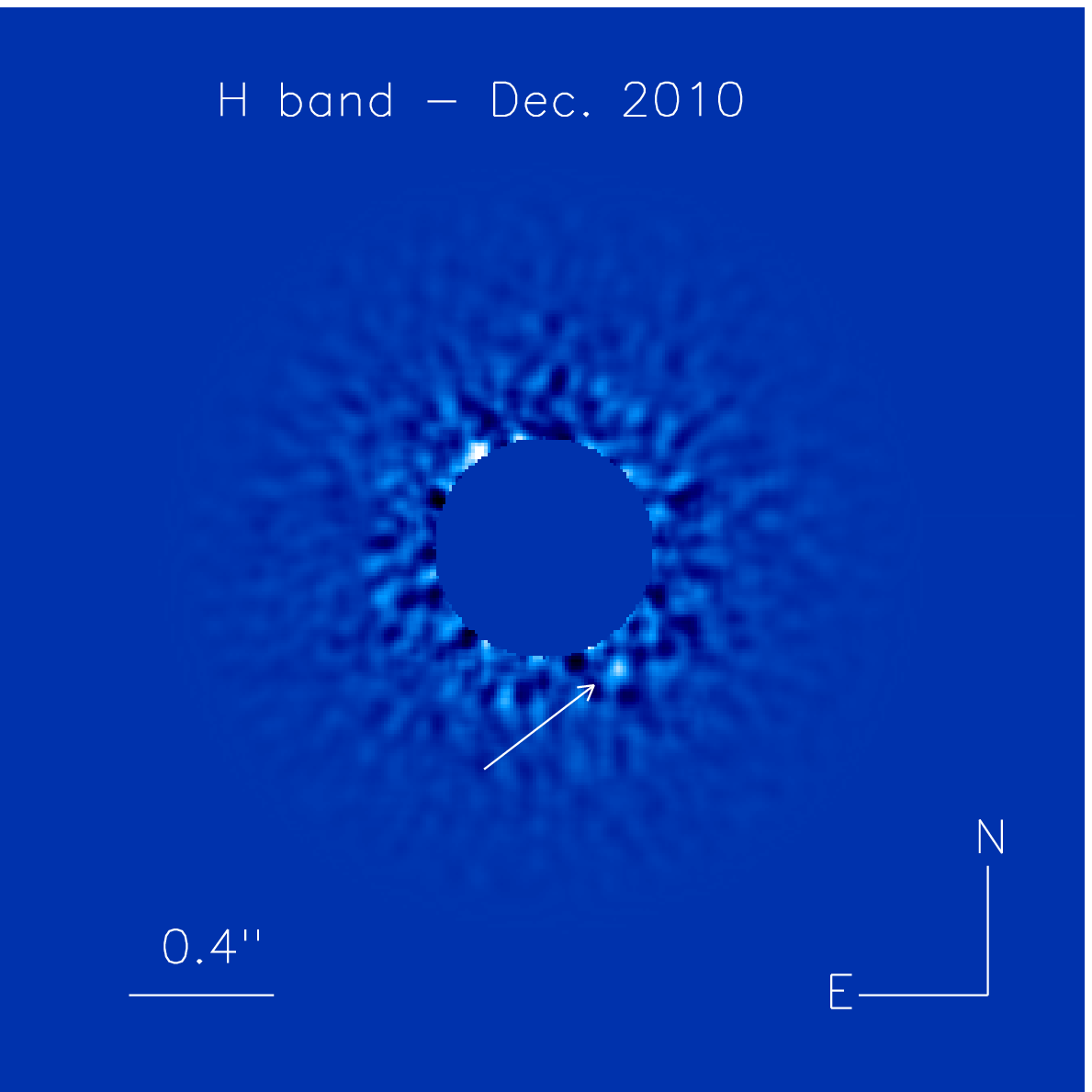}}

\caption{Final images processed with rADI (left) and LOCI (right) for the Ks and H filter (set 6). The arrow points to the planet. The central numerical mask is 0.3" in radius so larger than the coronagraphic mask. The field of view is 3"$\times$3".}
\label{fig:images}
\end{figure}

As a result, the data taken in 2008 do not show any hints of $\beta$ Pictoris b as expected from NaCo observations \citep{Lagrange2009b} the planet being angularly too close to the star (about 0.2" in projected separation  according to the orbit determination in \citet{Chauvin2012}). For the 2009 set of data (CH4L/CH4S and H), it is yet undetected since the Lyot mask is too large (r=0.32") with respect to the planet separation ($\rho=0.3"$ in Dec. 2009) and/or the halo is saturated (set 4). We note that it was recovered to the SW of the star in NaCo maskless images for the same epoch (Oct. to Dec. 2009) but at longer wavelengths (Lp filter, 3.8$\mu$m) thanks to a lower brightness ratio \citep{Lagrange2010}. 
 Moreover, the planet is too warm for methane to produce an absorption \citep[T=1700\,K,][]{Bonnefoy2013} so spectral differential imaging using CH4L/CH4S filters would not help to detect it.
Finally, the planet $\beta$ Pictoris b is detected in the Dec. 2010 images (sets 5 and 6) in the Ks band at SNR$ = 11-15$ (measured on a 3-pixel radius aperture with rADI and LOCI) although the individual frames are near the {  saturation level 
within 0.35" } from the mask (for the shortest integration time of 0.76s). The planet is also seen marginally in the H band (SNR$ = 5-6$). 
Figure \ref{fig:images} presents some images obtained with the rADI and LOCI algorithms for both the Ks and H bands. The results obtained with KLIP are not particularly of good quality (not shown here) and are actually strongly affected by the saturation of the frames near the edges of the mask.

\section{Photometry and Astrometry}

\begin{table*}[t]
\caption{Signal to noise ratio, magnitude differences in Ks (set 6) and astrometry ( $\rho$, PA), measured at the planet location for several algorithms.}
\begin{center}
\begin{tabular}{lccccc} \hline \hline
Algorithm 		&	SNR		&	FPOS			& 	FPNEG 				& $\rho$			& PA			   \\
			&			&	(mag)			&	(mag)				& (mas)			&	($\degb$)	 \\ \hline
rADI 0.50		&	13.3		&	$ {8.71\pm 0.49}$	& 	$ {8.32 	\pm 0.58}$	& $417 \pm 10$	& $210.98 \pm 0.70  $ \\	
rADI 1.00		&	15.6		&	$ {8.50\pm 0.49}$	& 	$ {8.57 	\pm 0.58}$	& $402 \pm 10$	&$212.30 \pm 0.70 $ \\
rADI 1.50		&	11.3		&	$ {8.72\pm 0.49}$	& 	$ {8.77 	\pm 0.58}$ 	& $402 \pm 10$	&$212.28  \pm 0.70  $ \\ 
LOCI 0.25		&	7.9		&	$ {8.82\pm 0.49}$	&   	$ {8.70 	\pm 0.66}$	&  $402 \pm 10$	&$212.30 \pm 0.70  $\\
LOCI 0.50		&	9.9		&	$ {9.06\pm 0.49}$	&   	$ {9.24 	\pm 0.66}$	&  $402\pm 10$ 	& $212.31 \pm 0.70  $\\
LOCI 0.75		&	8.2		&	$ {8.95\pm 0.49}$	&  	$ {9.25 	\pm 0.66}$	&  $401\pm 10$ 	&$212.44 \pm 0.70 $
\end{tabular}
\end{center}
\label{tab:photom}
\end{table*}%

In this section we focus on the analysis of set 6 which provides the best contrast performance at the location of the planet. The extraction of the planet photometry (and its error bar) is delicate in these data for two reasons, unsaturated PSFs are missing and the focal Lyot mask produces an attenuation on the planet flux.  

\begin{figure}[t]
\centerline{\includegraphics[width=8.5cm]{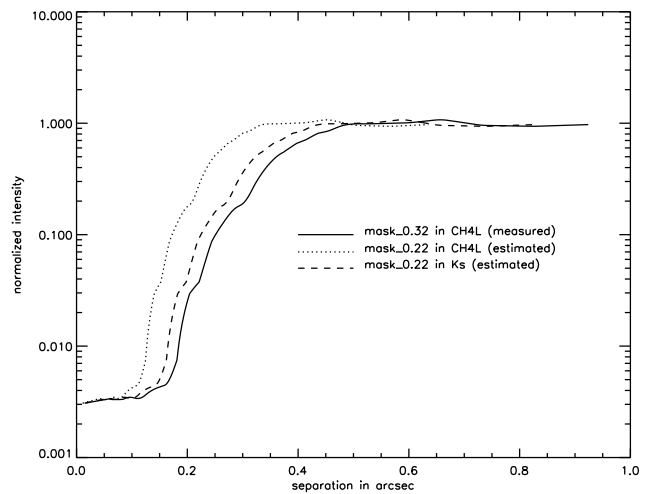}}
\caption{Off-axis transmissions of the mask derived from the 0.32" mask profile measured in CH4L (solid line) and estimated for the 0.22"/CH4L (dotted) and the Ks filter (dashed).}
\label{fig:maskTR}
\end{figure}

There is no published or reported calibration for the 0.22" mask attenuation profile. But, according to  the NICI instrument description, it is expected to be a scaled-down version of the 0.32" mask attenuation. We retrieved measurements obtained in CH4L for the 0.32" mask transmission as a function of X,Y coordinates. First, we derived the mask center coordinates from this low resolution profile, averaged the left and right parts and normalized to the intensity in the 0.5-0.8" region where the mask is expected to have no influence. Then, we assumed a proportional scaling with respect to the size of the mask (0.22/0.32) as well as the wavelength (CH4L/Ks = 2.15/1.652), the mask appearing like if it were smaller at longer wavelengths. The profiles we derived are shown in Fig. \ref{fig:maskTR}. 
Assuming the separation was $390\pm10$\,mas \citep{Chauvin2012} in Dec. 2010, we measured that the radial attenuation reaches $0.80\pm0.02$ at the location of $\beta$ Pictoris b.
As for the central attenuation of a point source, is no longer scalable to other wavelengths or mask sizes because of the diffraction by the coronagraphic mask.
Instead, to measure this central attenuation, we used a binary star observed in Ks in and out of the 0.22" mask (same object as in \citet{Biller2010}) and found a value of $120\pm 5$.
We retained this calibration to derive the planet photometry and we note that his number is very different from the  attenuation given in the NICI documentation (about 300). In addition, it also depends on the star centering and AO correction which are not necessarily identical when observing the target star. Such variations make the use of the central coronagraphic spot a poor photometric calibrator.
Overall, the uncertainties on the determination of the mask  radial transmission  and central attenuation contributes at a level of 8-10\% so 0.14\,mag hence comparable to the estimation of 11\% by \citet{Biller2010} for PZ Tel b with the same mask/filter configuration (PZ Tel b being at about the same separation as  $\beta$ Pictoris b).

To extract the photometry of the planet we proceeded in the same way as in \citet{Bonnefoy2011, Bonnefoy2013} so that it is directly comparable to the values derived with NaCo. Fake planets (attenuated copy of the PSF) were injected at the same radius as the planet ($\sim$0.4") but different PAs so that the speckle noise is averaged. We considered seven positive fake planets (FPPOS) equally separated by 45\deg and one negative fake planet (FPNEG) superimposed with $\beta$ Pictoris b. We measured the average intensity ratio between positive fake planets and the true planet (RPOS) as well as the residual intensity at the true planet location when using a negative fake planet (RNEG). As we scanned the fake planet to star brightness ratio from $2.10^{-5}$ to $2.10^{-4}$, we search for both the value of RPOS closer to unity and the minimal value of RNEG. 
The method was repeated for several rADI and LOCI reductions and the measured brightness ratios are reported in Tab. \ref{tab:photom} together with the SNRs. 
As for the determination of uncertainties, in addition to the errors introduced by the un-perfect knowledge on the mask transmission (0.14\,mag), we also took into account the observed PSF photometric variations (0.23\,mag) as well as the dispersion of our measurements (0.12\,mag for rADI/FPPOS, 0.21\,mag for rADI/FPNEG, 0.12\,mag for LOCI/FPPOS and 0.29\,mag for LOCI/FPNEG). To be conservative the errors were added linearly amounting to the final error bars indicated in Tab. \ref{tab:photom}.

Therefore, the derived magnitude differences of Tab. \ref{tab:photom} are only in marginal agreement  with respect to those measured with NaCo ($\Delta m_{Ks}=9.2\pm0.2$). 
While this is compatible within the error bars we attribute the discrepancy  to 1/ the lack of an unocculted, unsaturated PSF for the target star, and 2/ from the saturation of the coronagraphic halo near the position of the planet.


As for the astrometry, we assumed the star position is determined by the central attenuated spot behind the coronagraphic mask and we performed Moffat fitting in a small 5-pixel radius aperture centered at the guessed planet location. The obtained values (Tab. \ref{tab:photom}): $\rho = 404 \pm 10$\,mas and $PA = 212.1 \pm 0.7 \degb$ on average, are in perfect agreement with those found by \citet{Chauvin2012} in Nov. 2010 ($390\pm10$\,mas, $PA=212.34\pm2.13\degb$), so one month earlier than the present data. The True North position is known at an accuracy better than $0.1\degb$ (included in our error bar) according to the NICI documentation.


\section{Conclusion}
The NICI data released on Nov. 24th, 2012 allowed us to recover the planet $\beta$ Pictoris b at $404\pm10$\,mas from the star in the South West direction from Dec. 2010 data. It is the first time $\beta$ Pictoris b is unambiguously detected with a telescope and an instrument which is 
different than the discovery instrument NaCo. Although the presence of the planet is already well attested from NaCo observations owing to careful follow-up, the present result 
provides 
an healthy independent confirmation and demonstrates the importance of archiving high contrast imaging data
 as already shown for the HR\,8799 planets  \citep{Lafreniere2009, Soummer2011, Currie2012}. 
The astrometry of the companion is in conformity with respect to NaCo observations. However, we note that the lack of a true PSF of $\beta$ Pic
prevents us to compare accurately the photometry with previous estimations from NaCo. 
More observations with NICI are desirable to collect additional photometric measurements and then further explore the atmospheric characterization of this planet.
With the upcoming next generation of planet finders and the hopefully large number of planetary mass companion detection, it will be mandatory to perform multi-telescope observations for independent confirmations.  

\begin{acknowledgements}
We would like to thank the anonymous referee for providing relevant comments which helped us to improve this letter. 
\end{acknowledgements}

\bibliography{bpicnici.bib}

\begin{thebibliography}{31}
\expandafter\ifx\csname natexlab\endcsname\relax\def\natexlab#1{#1}\fi

\bibitem[{Baraffe {et~al.}(2003)Baraffe, Chabrier, Barman, Allard, \&
  Hauschildt}]{Baraffe2003}
Baraffe, I., Chabrier, G., Barman, T.~S., Allard, F., \& Hauschildt, P.~H.
  2003, Astronomy {\&} Astrophysics, 402, 701

\bibitem[{{Beuzit} {et~al.}(2008){Beuzit}, {Feldt}, {Dohlen}, {Mouillet},
  {Puget}, {Wildi}, {Abe}, {Antichi}, {Baruffolo}, {Baudoz}, {Boccaletti},
  {Carbillet}, {Charton}, {Claudi}, {Downing}, {Fabron}, {Feautrier},
  {Fedrigo}, {Fusco}, {Gach}, {Gratton}, {Henning}, {Hubin}, {Joos}, {Kasper},
  {Langlois}, {Lenzen}, {Moutou}, {Pavlov}, {Petit}, {Pragt}, {Rabou}, {Rigal},
  {Roelfsema}, {Rousset}, {Saisse}, {Schmid}, {Stadler}, {Thalmann}, {Turatto},
  {Udry}, {Vakili}, \& {Waters}}]{Beuzit2008}
{Beuzit}, J.-L., {Feldt}, M., {Dohlen}, K., {et~al.} 2008, in Society of
  Photo-Optical Instrumentation Engineers (SPIE) Conference Series, Vol. 7014,
  Society of Photo-Optical Instrumentation Engineers (SPIE) Conference Series

\bibitem[{Biller {et~al.}(2010)Biller, Liu, Wahhaj, Nielsen, Close, Dupuy,
  Hayward, Burrows, Chun, Ftaclas, Clarke, Hartung, Males, Reid, Shkolnik,
  Skemer, Tecza, Thatte, Alencar, Artymowicz, Boss, de~Gouveia Dal~Pino,
  Gregorio-Hetem, Ida, Kuchner, Lin, \& Toomey}]{Biller2010}
Biller, B.~A., Liu, M.~C., Wahhaj, Z., {et~al.} 2010, The Astrophysical Journal
  Letters, 720, L82

\bibitem[{Boccaletti {et~al.}(2012)Boccaletti, Augereau, Lagrange, Milli,
  Baudoz, Mawet, Mouillet, Lebreton, \& Maire}]{Boccaletti2012}
Boccaletti, A., Augereau, J.~C., Lagrange, A.~M., {et~al.} 2012, Astronomy {\&}
  Astrophysics, 544, 85

\bibitem[{{Bonnefoy} {et~al.}(2013){Bonnefoy}, {Boccaletti}, {Lagrange},
  {Allard}, {Mordasini}, {Beust}, {Chauvin}, {Girard}, {Homeier}, {Apai},
  {Lacour}, \& {Rouan}}]{Bonnefoy2013}
{Bonnefoy}, M., {Boccaletti}, A., {Lagrange}, A.-M., {et~al.} 2013, ArXiv
  e-prints

\bibitem[{Bonnefoy {et~al.}(2011)Bonnefoy, Lagrange, Boccaletti, Chauvin, Apai,
  Allard, Ehrenreich, Girard, Mouillet, Rouan, Gratadour, \&
  Kasper}]{Bonnefoy2011}
Bonnefoy, M., Lagrange, A.~M., Boccaletti, A., {et~al.} 2011, Astronomy {\&}
  Astrophysics, 528, L15

\bibitem[{Chauvin {et~al.}(2012)Chauvin, Lagrange, Beust, Bonnefoy, Boccaletti,
  Apai, Allard, Ehrenreich, Girard, Mouillet, \& Rouan}]{Chauvin2012}
Chauvin, G., Lagrange, A.~M., Beust, H., {et~al.} 2012, Astronomy {\&}
  Astrophysics, 542, 41

\bibitem[{Currie {et~al.}(2012)Currie, Fukagawa, Thalmann, Matsumura, \&
  Plavchan}]{Currie2012}
Currie, T., Fukagawa, M., Thalmann, C., Matsumura, S., \& Plavchan, P. 2012,
  arXiv.org, astro-ph.EP

\bibitem[{Fitzgerald {et~al.}(2009)Fitzgerald, Kalas, \&
  Graham}]{Fitzgerald2009}
Fitzgerald, M.~P., Kalas, P.~G., \& Graham, J.~R. 2009, The Astrophysical
  Journal Letters, 706, L41

\bibitem[{Heap {et~al.}(2000)Heap, Lindler, Lanz, Cornett, Hubeny, Maran, \&
  Woodgate}]{Heap2000}
Heap, S.~R., Lindler, D.~J., Lanz, T.~M., {et~al.} 2000, The Astrophysical
  Journal, 539, 435

\bibitem[{Lafreni{\`e}re {et~al.}(2009)Lafreni{\`e}re, Marois, Doyon, \&
  Barman}]{Lafreniere2009}
Lafreni{\`e}re, D., Marois, C., Doyon, R., \& Barman, T. 2009, The
  Astrophysical Journal Letters, 694, L148

\bibitem[{{Lafreni{\`e}re} {et~al.}(2007){Lafreni{\`e}re}, {Marois}, {Doyon},
  {Nadeau}, \& {Artigau}}]{Lafreniere2007}
{Lafreni{\`e}re}, D., {Marois}, C., {Doyon}, R., {Nadeau}, D., \& {Artigau},
  {\'E}. 2007, \apj, 660, 770

\bibitem[{{Lagrange} {et~al.}(2012){Lagrange}, {Boccaletti}, {Milli},
  {Chauvin}, {Bonnefoy}, {Mouillet}, {Augereau}, {Girard}, {Lacour}, \&
  {Apai}}]{Lagrange2012b}
{Lagrange}, A.-M., {Boccaletti}, A., {Milli}, J., {et~al.} 2012, \aap, 542, A40

\bibitem[{{Lagrange} {et~al.}(2010){Lagrange}, {Bonnefoy}, {Chauvin}, {Apai},
  {Ehrenreich}, {Boccaletti}, {Gratadour}, {Rouan}, {Mouillet}, {Lacour}, \&
  {Kasper}}]{Lagrange2010}
{Lagrange}, A.-M., {Bonnefoy}, M., {Chauvin}, G., {et~al.} 2010, Science, 329,
  57

\bibitem[{Lagrange {et~al.}(2012)Lagrange, De~Bondt, Meunier, Sterzik, Beust,
  \& Galland}]{Lagrange2012a}
Lagrange, A.~M., De~Bondt, K., Meunier, N., {et~al.} 2012, Astronomy {\&}
  Astrophysics, 542, 18

\bibitem[{Lagrange {et~al.}(2009{\natexlab{a}})Lagrange, Gratadour, Chauvin,
  Fusco, Ehrenreich, Mouillet, Rousset, Rouan, Allard, Gendron, Charton,
  Mugnier, Rabou, Montri, \& Lacombe}]{Lagrange2009a}
Lagrange, A.~M., Gratadour, D., Chauvin, G., {et~al.} 2009{\natexlab{a}},
  Astronomy {\&} Astrophysics, 493, L21

\bibitem[{Lagrange {et~al.}(2009{\natexlab{b}})Lagrange, Kasper, Boccaletti,
  Chauvin, Gratadour, Fusco, Ehrenreich, Apai, Mouillet, \&
  Rouan}]{Lagrange2009b}
Lagrange, A.~M., Kasper, M., Boccaletti, A., {et~al.} 2009{\natexlab{b}},
  Astronomy {\&} Astrophysics, 506, 927

\bibitem[{{Lenzen} {et~al.}(2003){Lenzen}, {Hartung}, {Brandner}, {Finger},
  {Hubin}, {Lacombe}, {Lagrange}, {Lehnert}, {Moorwood}, \&
  {Mouillet}}]{Lenzen2003}
{Lenzen}, R., {Hartung}, M., {Brandner}, W., {et~al.} 2003, in Society of
  Photo-Optical Instrumentation Engineers (SPIE) Conference Series, Vol. 4841,
  Society of Photo-Optical Instrumentation Engineers (SPIE) Conference Series,
  ed. {M.~Iye \& A.~F.~M.~Moorwood}, 944--952

\bibitem[{Liu {et~al.}(2010)Liu, Wahhaj, Biller, Nielsen, Chun, Close, Ftaclas,
  Hartung, Hayward, Clarke, Reid, Shkolnik, Tecza, Thatte, Alencar, Artymowicz,
  Boss, Burrows, de~Gouveia Dal~Pino, Gregorio-Hetem, Ida, Kuchner, Lin, \&
  Toomey}]{Liu2010}
Liu, M.~C., Wahhaj, Z., Biller, B.~A., {et~al.} 2010, Adaptive Optics Systems
  II. Edited by Ellerbroek, 7736, 53

\bibitem[{Lloyd \& Sivaramakrishnan(2005)}]{Lloyd2005}
Lloyd, J.~P. \& Sivaramakrishnan, A. 2005, The Astrophysical Journal, 621, 1153

\bibitem[{{Macintosh} {et~al.}(2008){Macintosh}, {Graham}, {Palmer}, {Doyon},
  {Dunn}, {Gavel}, {Larkin}, {Oppenheimer}, {Saddlemyer}, {Sivaramakrishnan},
  {Wallace}, {Bauman}, {Erickson}, {Marois}, {Poyneer}, \&
  {Soummer}}]{Macintosh2008}
{Macintosh}, B.~A., {Graham}, J.~R., {Palmer}, D.~W., {et~al.} 2008, in Society
  of Photo-Optical Instrumentation Engineers (SPIE) Conference Series, Vol.
  7015, Society of Photo-Optical Instrumentation Engineers (SPIE) Conference
  Series

\bibitem[{{Marois} {et~al.}(2006){Marois}, {Lafreni{\`e}re}, {Doyon},
  {Macintosh}, \& {Nadeau}}]{Marois2006}
{Marois}, C., {Lafreni{\`e}re}, D., {Doyon}, R., {Macintosh}, B., \& {Nadeau},
  D. 2006, \apj, 641, 556

\bibitem[{Mouillet {et~al.}(1997)Mouillet, Larwood, Papaloizou, \&
  Lagrange}]{Mouillet1997}
Mouillet, D., Larwood, J.~D., Papaloizou, J. C.~B., \& Lagrange, A.~M. 1997,
  Monthly Notices of the Royal Astronomical Society, 292, 896

\bibitem[{Nielsen {et~al.}(2012)Nielsen, Liu, Wahhaj, Biller, Hayward, Boss,
  Bowler, Kraus, Shkolnik, Tecza, Chun, Clarke, Close, Ftaclas, Hartung, Males,
  Reid, Skemer, Alencar, Burrows, de~Gouveia Dal~Pino, Gregorio-Hetem, Kuchner,
  Thatte, \& Toomey}]{Nielsen2012}
Nielsen, E.~L., Liu, M.~C., Wahhaj, Z., {et~al.} 2012, The Astrophysical
  Journal, 750, 53

\bibitem[{Quanz {et~al.}(2010)Quanz, Meyer, Kenworthy, Girard, Kasper,
  Lagrange, Apai, Boccaletti, Bonnefoy, Chauvin, Hinz, \& Lenzen}]{Quanz2010}
Quanz, S.~P., Meyer, M.~R., Kenworthy, M.~A., {et~al.} 2010, The Astrophysical
  Journal Letters, 722, L49

\bibitem[{Racine {et~al.}(1999)Racine, Walker, Nadeau, Doyon, \&
  Marois}]{Racine1999}
Racine, R., Walker, G. A.~H., Nadeau, D., Doyon, R., \& Marois, C. 1999, The
  Publications of the Astronomical Society of the Pacific, 111, 587

\bibitem[{{Rousset} {et~al.}(2003){Rousset}, {Lacombe}, {Puget}, {Hubin},
  {Gendron}, {Fusco}, {Arsenault}, {Charton}, {Feautrier}, {Gigan}, {Kern},
  {Lagrange}, {Madec}, {Mouillet}, {Rabaud}, {Rabou}, {Stadler}, \&
  {Zins}}]{Rousset2003}
{Rousset}, G., {Lacombe}, F., {Puget}, P., {et~al.} 2003, in Society of
  Photo-Optical Instrumentation Engineers (SPIE) Conference Series, Vol. 4839,
  Society of Photo-Optical Instrumentation Engineers (SPIE) Conference Series,
  ed. {P.~L.~Wizinowich \& D.~Bonaccini}, 140--149

\bibitem[{Soummer {et~al.}(2011)Soummer, Brendan~Hagan, Pueyo, Thormann, Rajan,
  \& Marois}]{Soummer2011}
Soummer, R., Brendan~Hagan, J., Pueyo, L., {et~al.} 2011, The Astrophysical
  Journal, 741, 55

\bibitem[{Soummer {et~al.}(2012)Soummer, Pueyo, \& Larkin}]{Soummer2012}
Soummer, R., Pueyo, L., \& Larkin, J. 2012, arXiv.org, astro-ph.IM

\bibitem[{Toomey \& Ftaclas(2003)}]{Toomey2003}
Toomey, D.~W. \& Ftaclas, C. 2003, Instrument Design and Performance for
  Optical/Infrared Ground-based Telescopes. Edited by Iye, 4841, 889

\bibitem[{Wahhaj {et~al.}(2011)Wahhaj, Liu, Biller, Clarke, Nielsen, Close,
  Hayward, Mamajek, Cushing, Dupuy, Tecza, Thatte, Chun, Ftaclas, Hartung,
  Reid, Shkolnik, Alencar, Artymowicz, Boss, de~Gouveia Dal~Pino,
  Gregorio-Hetem, Ida, Kuchner, Lin, \& Toomey}]{Wahhaj2011}
Wahhaj, Z., Liu, M.~C., Biller, B.~A., {et~al.} 2011, The Astrophysical
  Journal, 729, 139

\end{thebibliography}
\end{document}